# Wilbur Norman Christiansen 1913-2007

*R. H. Frater[A] and W. M. Goss[B]*

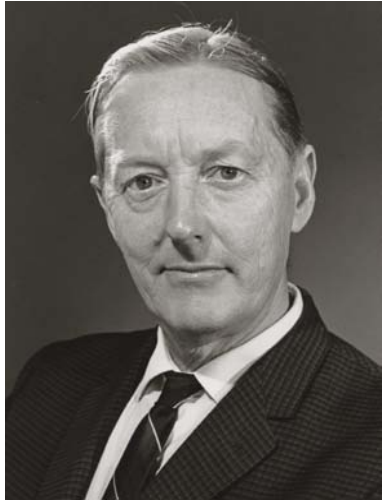

W. N. ('Chris') Christiansen was an innovative and influential radio astronomy pioneer. The hallmarks of his long and distinguished career in science and engineering, spanning almost five decades, were his inventiveness and his commitment to, and success with, large-scale projects. These projects were the outcome of his innovative skill as physicist and engineer. Paralleling this was his equal commitment to forging strong international links and friendships, leading to his election as Vice-President of the International Astronomical Union for the years 1964 to 1970, as President of the International Union of Radio Science, URSI, from 1978 to 1981, and subsequently as Honorary Life President in 1984, and as Foreign Secretary of the Australian Academy of Science from 1981 to 1985. Major subsequent developments in radio astronomy and wireless communications on the global scene stand as a legacy to Chris's approach to his work and to the development of those who worked with him.

## Early Years

Wilbur Norman Christiansen ('Chris' to everyone) was born on 9 August 1913 in the Melbourne suburb of Elsternwick, where his father was the minister of the local Congregational Church.

Chris's grandfather, Jens Christiansen, came from Denmark as a teenager, jumping ship in Melbourne and walking to the goldfields. Chris's father was one of eight children. He went to the one-teacher school at Three Mile and later to Beechworth College, in the north-east Victorian goldfields. He earned money around the goldfields by tutoring. Later he returned to Melbourne, worked at the Public Library and did a university course. He later studied theology and became a clergyman.

When Chris was a small child, his father was posted to Perth and Chris started school at the local State primary school there. His father died of peritonitis at the age of 37 in the early 1920s. After his father's death, the family returned to Melbourne, where Chris's mother taught music to support the family and Chris went to the local primary school. Chris spoke of the environment he grew up in:

> Strangely enough, having spent all her time looking after her brats and teaching music, she seemed to have nothing to do in the evening but to play the piano, and being the last to go to sleep (because I was the eldest) I always went to sleep listening to Beethoven, Brahms and so on.

Chris subsequently went to Caulfield Grammar School where the headmaster had offered the family free tuition.

[A] Corresponding author. PO Box 456, Lindfield, NSW 2070, Australia. E-mail: Bob.Frater@resmed.com.au

[B] National Radio Astronomy Observatory, PO Box 0, Socorro, New Mexico 87801, USA.

Chris was a keen hobbyist and an inventor from an early stage. He became secretary of both the school's camera club and the radio club. He proudly described an early invention, as a schoolboy, that allowed one to write five lines at a time when given 100 lines as a punishment.

He became knowledgeable about crystal sets. The crystal receiver consisted only of a capacitor, an inductance, a galena crystal rectifier, a pair of headphones and a cat's whisker to make the connection. Chris experimented with home-made lead sulphide because a piece of galena was too expensive. He talked, as well, of trying to make an amplifier with multiple cat's whiskers in the pre-transistor era.

## The University of Melbourne – Science, Philosophy and Politics

Chris started at the University of Melbourne in 1930, intending to study Architecture but at the last minute choosing Science. He spoke of being distracted from his studies by time spent in the library reading philosophy rather than physics, but in the end he persisted with physics. He described his reactions to the distress he saw during the great depression as follows:

> As I walked to the University each morning through a rather poor district I saw people and their furniture being ejected from their houses. My early Christian beliefs that had not been buried by my now complete unbelief in the supernatural made me react very strongly. I joined the University Labour Club, which was very much to the left of the Labour Party, and became very active in this. We published a journal called Proletariat and also a weekly news sheet and joined marches of the unemployed and were ridden down by the mounted police. At the University there were some supporters of Mussolini and Hitler and these tried to break up our meetings until the V.C. [Vice-Chancellor] stopped them. In addition to the Labour Club activities I went to any other lunch time meeting or musical recital that was taking place and after lectures I joined the swimming club, I played hockey in one of the University teams and when the 'lefties' were being attacked by the 'fascisti' I joined the boxing club.

It was at a weekend camp of the Melbourne University Labor Club that Chris met his wife-to-be Elsie (later Elspeth).

Chris graduated BSc in 1934 and MSc with First Class Honours in 1935 from the University of Melbourne. He was awarded the Dwight Prize in Physics, which he said paid for his books, in 1931, and the Kernot Prize in 1934.

While a postgraduate student at Melbourne in 1935 he discovered, with Crabtree and Laby, that 'light' and 'heavy' water could be separated by fractional distillation, indicating that special measures must be taken in purifying water prior to analysing its isotopic content (1). Laby was interested in building up a supply of deuterium for his planned programme of nuclear research.

Chris described aspects of this work and a dispute with Professor Laby:

> Heavy water ($D_2O$) had recently been discovered and it was already known that deuterium could be separated from hydrogen by fractional electrolysis. For that reason old alkaline batteries that had been charged for many years had an electrolyte rich in deuterium. Such was available in cable-tram batteries in Melbourne and I was given the job of collecting supplies of this electrolyte and by further electrolysis producing water with a very high component of heavy water. A second part of the work was to produce water almost free of deuterium and by a very sensitive density measurement to find the proportion of deuterium in natural water. This work was highly dangerous. All the gases from the electrolysis of water were collected together and forced through a nozzle and the jet set alight. If the nozzle velocity fell below the burning velocity the flame could go backwards into the electrolysis chamber and a huge explosion might occur.

> Professor Laby had designed a device which allegedly reduced the size of the nozzle if the pressure fell and kept the jet - velocity high enough for safety and avoided having to replace one blown up research student by another one. I was suspicious of this and worked out the theory and found it would not work. I gave my calculations to the professor who angrily tossed them away and suggested that I might take up some other branch of scientific activity. I meekly said that I would try his device. I would attach it to an inflated balloon filled with the Oxy-hydrogen mixture and light the jet. As the gas escaped and was burned the pressure would fall and we would see how the safety device worked. I set this up in the basement of the Physics school, lit the jet and ran upstairs. In a few minutes there was an enormous bang that caused great excitement throughout the building.

I was then allowed to use my own safety device which was simply a jar of sand between the jet and the source of gas. I got a prize for my MSc work rather than either death or the sack. The success of my thesis was not due to my heavy water production about which Laby told Tom Cherry (the professor of mathematics) 'young Christiansen is going to blow himself up', but was the result of my finding that minute changes in the proportions of hydrogen and deuterium in common water could be produced by fractional distillation. Laby as an authority on physical standards was delighted that water density could no longer be used as a standard. 'Well done, young Christiansen!'.

Chris's research career was on its way. In 1953, by which time he was working with CSIRO, he submitted his collected papers of the time and was awarded a Doctor of Science degree by the University of Melbourne.

## Early Working Life

In 1937, after completing a two-year appointment as Assistant Physicist with the Commonwealth X-ray and Radium Laboratory, Chris got a job at Amalgamated Wireless (Australasia) Ltd (AWA) in Sydney. Elspeth moved from Melbourne to teach at Ravenswood School, in Sydney, and they got married.

At that stage, AWA manufactured radios and communications equipment, and operated the 'Beam Wireless' system for Australia's overseas telecommunications. The company was half-owned by the Australian Government. The Beam Wireless section would ultimately become the Overseas Telecommunications Commission (OTC).

Chris joined Geoffrey Builder and A. L. Green, pioneers in ionospheric physics research, at AWA's research laboratory in Sydney, where his major work was concentrated on improving the 'Beam Wireless' system and particularly on the design of stacked rhombic antennas for overseas short-wave communications. He made an important contribution to securing Australia's international wartime radio communication linkages, and his experience in this area was also valuable in his later roles in CSIR (2, 3, 4, 5, 6, 7).

Published in the *AWA Technical Review*, this work appeared subsequently in the CCIR (International Radio Consultative Committee) 'High Frequency Directional Antennae' handbook and was widely referenced. Chris often told us: 'I got five bob for my invention'. The Overseas Telecommunications Commission, AWA's postwar successor in operating the short-wave services, made extensive use of the designs. As a young graduate at OTC in 1960, one of us (RHF) found himself designing rhombic antennas using the graphs and tables produced by Chris years earlier.

This period of Chris's career introduced him to a wide range of technologies and to some of the best technical people of the time, including Geoffrey Builder, A. L.Green, Ernie Benson, and Ruby Payne-Scott and Lindsay McCready who like him would go on to Radiophysics. Although both Chris and Ruby complained about AWA, there were many learning opportunities that a person with a 'hands-on' approach like Chris's was able to take up and carry with him into his later life.

Chris was very proud of his 'ecumenical' role, in the 1940s, in arranging to have radio stations 2CH (the Protestant station) and 2SM (the Roman Catholic station) broadcast from the one antenna at Carlingford, an outer suburb of Sydney. He was able to apply the same approach of sharing antennas between stations in a number of locations in New Zealand, helping AWA's position in that market considerably.

## The Radiophysics Division of CSIR/CSIRO

Chris had a long-standing interest in astronomy and, in 1948, wrote to the Radiophysics Division of the Council of Scientific and Industrial Research (CSIR, later CSIRO), enquiring about positions. The Division was then headed by E. G. ('Taffy') Bowen, and a group headed by J. L. Pawsey had been doing pioneering work in radio astronomy. Chris was offered a position in Pawsey's group.

Chris enjoyed the benefits of joining an environment that been fostered by CSIR's longstanding Chief Executive Officer, David Rivett: 'get the best people possible, give them the needed resources and let them run free'. As Chris pointed out, in practice resources were never plentiful, but the freedom and encouragement was there. In this regard, Joe Pawsey was a man in the Rivett mould and he earned enormous respect from all those who worked with him. Chris was a great admirer and in his own later roles emulated Pawsey's approach.

Chris was appointed to a senior role within Radiophysics, filling a vacancy created by the transfer of Fred Lehany to CSIR's Division of Electrotechnology. He was soon to be the leader of the solar research programme at the newly-established field station at Potts Hill in the western suburbs of Sydney. The main radio telescope there was a 16 x 18-ft wartime experimental radar that had been relocated from the Georges Heights field station to Potts Hill in time for the 1948 solar eclipse (see Wendt et al., 2008a). Chris organized observations of partial solar eclipses with D. E. Yabsley and B. Y. Mills at a wavelength of 50 cm in 1948 and with Yabsley at a wavelength of 25 cm in 1949. These observations showed that regions of the Sun of high radio emission (associated with sunspots) had dimensions of about one-tenth of the solar diameter. They were the subject of Chris's first astronomy publication (9).

This discovery and Chris's frustration at the inefficiency of depending on eclipses for measurements led to the development of the 'grating array' that achieved high resolution as a result of its length and produced multiple responses on the sky separated by a number of solar diameters. The first grating telescope at Potts Hill (1951) allowed the distribution of radio brightness across the Sun to be studied as the sun drifted through the responses during the day. This telescope array was a very manual affair, requiring the observers to run up and down the array adjusting the pointing of the individual antennas. It was found that the enhanced emission at 21cm came from regions in the lower corona of the Sun. Their dimensions and height above the photosphere could be determined for the first time.

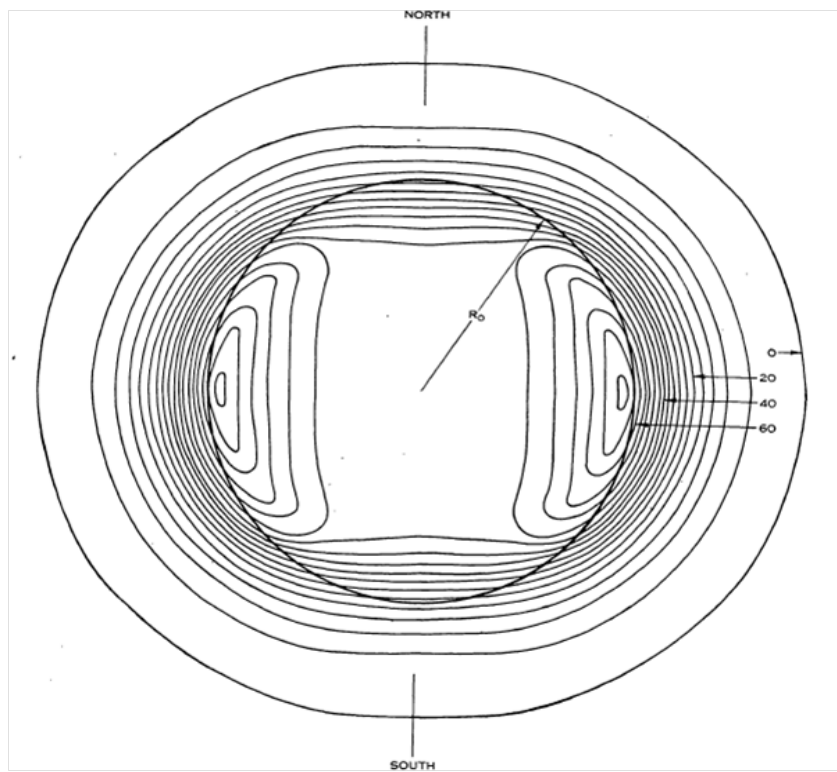

Figure 2. The First earth-rotational synthesis image showing the derived two-dimensional
radio brightness distribution with a peak Tb of 6.8e4 K and the central value of 4.7e4K

A second investigation concerned the background or 'quiet sun' radiation at 21 cm. David Martyn had predicted 'limb brightening' but the Cambridge radio astronomy group had conducted observations that discounted the possibility. Chris, with J. A. Warburton, observed limb brightening with early experiments using an east-west array. They later made use of two gratings, one in an east-west and one in a north-south direction, to observe the Sun from dawn until dusk to obtain the two-dimensional brightness distribution. This showed marked limb brightening in the equatorial zones but none at the poles. The method that was used for this showed a lot of ingenuity and deserves discussion here. The scans produced as the sun drifted through each grating response (one-dimensional pictures of the Sun known as 'strip scans') were Fourier-transformed to produce a radial line of points with an azimuth angle equal to the scanning angle across the Sun. (Because the observed scan distributions were symmetrical, they only computed the cosine transform.) This angle changed during the day with the rotation of the Earth.  At the end of this stage of processing, they had sets of points on radial lines covering a good part of a circular patch in the centre of the Fourier plane. One of the issues at this stage is that the density of points is inversely proportional to the distance from the centre, giving undue weight to broader features and low weight to finer features. This could be corrected by weighting the points on each line.  The approach they used to form the final image involved producing strip scans across this set of radial data points. They simply summed the values under a strip moving at a fixed angle across the data.  This was again Fourier-transformed, resulting in a radial line back in the 'image' plane.  The set of radial distributions was then used to create a two-dimensional contour map. This process was carried out at enough angles to give the detail required in plotting contours for the final image.  This involved laborious hand-calculation and took months of effort to produce an image, shown in Figure 2.

This approach was the first known application of earth-rotational synthesis and produced for the first time a radio map with a resolution as fine as four minutes of arc.

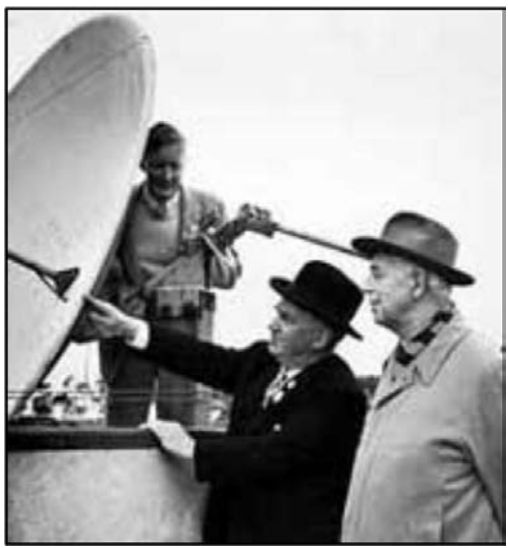

Figure 3.  Chris is demonstrating part of the telescope at Pott's Hill to Professor B. van der Pol,

the famous Dutch engineer and scientist. There also, with his black hat on, is Sir Edward Appleton.

This image was taken during the URSI General Assembly in Sydney in 1952.

One of the people who worked with Chris at this time was Govind Swarup, who had come to Australia under the Colombo Plan from the National Physical Laboratory in India.  As he said in a recent paper, 'Reminiscences regarding Professor W. N. Christiansen': 'I learned the powerful technique of radio interferometry from Chris in 1953 and I haven't looked back'.

When the grating array at Potts Hill was no longer being used, Pawsey and Chris arranged for it to be transferred to India where it ultimately became the Kalyan Radio Telescope and began observing in 1965.

## The Hydrogen Line

In 1951, news came from E. M. Purcell in America that one of his students, H. I. Ewen, after working for a couple of years, had found a particular radio spectral line from space that would make it possible to look not at the hot or active parts of the interstellar medium but at the hydrogen in the cold regions.   In contrast to optical wavelengths, there was no absorption due to the pervasive galactic dust. This important discovery had been predicted by Hendrik van de Hulst in Holland. Purcell sent requests to Joe Pawsey and the Dutch groups to help confirm the discovery.

Chris worked with J. V. Hindman on this project:

> All we knew was that this had been seen with a fixed aerial when the Milky Way went through. With Jim Hindman as my assistant I got stuck into this very rapidly, using all sorts of old junk that we could collect. We did have to make up one special bit of instrumentation. Within six weeks we confirmed that the radiation was coming from the Milky Way, but we went further and mapped it all over the Milky Way, showing that in fact it had exactly the same shape as the Milky Way. Moreover, because we were doing it in a special way, we were able to show there were spiral arms in our galaxy – the first radio evidence that we were living in a spiral galaxy.

As Chris remarked later:

Our research was done crudely but it was good fun and the results were exciting. When Purcell's research student Ewen came over (to Sydney for the 1952 URSI meeting) and saw the gear I had, with cables lying all over the floor and ancient oscillators, he said, 'My God, I can understand why you could do it in six weeks and it took me two years!'

Chris and Jim Hindman continued with their crude equipment. They had not only confirmed the discovery of 21 cm radiation from ground-state hydrogen in space but had made the first survey of H-line emission from space, obtaining the first radio evidence of the existence of spiral arms in our Galaxy, the Milky Way (12, 13).

Chris had had some remarkable opportunities and delivered excellent outcomes. He described the approach of Taffy Bowen and Joe Pawsey in the Division:

Pawsey's scientific style set the tone completely, but he was a very unworldly fellow. Bowen was the man who got the money, the tough businessman, while Joe was the rather academic scientist. And it was an excellent combination.

Joe Pawsey had a remarkable ability to distribute the work in ways that allowed a number of individuals to develop their talents.

## A French Interlude, then Fleurs and the 'Chris Cross'

The French were very interested indeed in the work at Radiophysics and, at the URSI General Assembly in Sydney in 1952, Chris was invited to go and work in France. He spent his time there at the Meudon Observatory, where he learned 'a lot about France and optical astronomy'. Back in the Radiophysics Laboratory, he started on a new project.

A new instrument, which combined the grating approach from Potts Hill with the Mills Cross principle conceived by Bernard Mills, was subsequently built at the Fleurs field station. Known as the 'Chris Cross', the telescope provided daily two-dimensional maps of the Sun with a resolution of 3 arc minutes from 1957 onwards (23). These maps were built up by scanning with a pencil beam to produce a 'raster scan' where the beam moved up one step at a time as the Sun went through successive grating lobes with the rotation of the earth. It was the first of a number of similar instruments built around the world.

During this period at CSIRO (1948-59), Chris was awarded the Syme Prize for Research by the University of Melbourne in 1959, and a paper describing the design of the Grating Cross received the Fleming Premium of the Institution of Electrical Engineers in 1961.

## A Parting of Ways

In the late 1950s, there were big arguments within Radiophysics on whether future needs should be met by array telescopes or by a big dish. Joe Pawsey, Chris and Bernard Mills favoured arrays while Taffy Bowen wanted a big dish. Pawsey, Chris and Mills all left, Chris and Mills going to the University of Sydney, to Chairs in Electrical Engineering and Physics respectively. Pawsey had a position in the USA with the National Radio Astronomy Observatory at Green Bank but died from a brain tumour before he could take it up. At the time, Chris referred to the future Parkes Telescope as the 'Last of the Windjammers'. The irony here is that Chris and Bernie Mills played key roles in finding the site near Parkes, New South Wales, where the dish was eventually built.

In a paper titled 'The Present Difficulties in Australian Radio Astronomy', dated 31 March 1960, Joe Pawsey set out his assessment of the state of radio astronomy in the era preceding the opening of the Parkes radio telescope on 31 October 1961. The paper was submitted to the CSIRO Executive in order to gain support for other new initiatives in Australia such as the proposed 'Super Cross ', an improved Mills Cross. Pawsey was also anxious to support Christiansen: 'The report recommends that CSIRO should be prepared to co-operate very closely with the new Professor of Electrical Engineering at Sydney University.' Pawsey pointed out that the curtailment of the solar work at Fleurs had led to a plan for the development of high-resolution techniques applicable to cosmic problems. The construction of a compound interferometer by adding an 18 metre paraboloid to the existing solar array of 5.8 metre antennas had been discussed. Pawsey concluded by stating that Christiansen wanted to co-operate with the CSIRO Division of Radiophysics: 'I believe this would be to our mutual advantage and consider our cooperation should take the form of a joint project.'

## Electrical Engineering at the University of Sydney and the Return to Fleurs

Chris was appointed to the Chair of Electrical Engineering at the University of Sydney in 1960. He did not take up his post immediately but first spent fifteen months at Leiden University, at the invitation of Professor Jan Oort, as the leader of an international design team for the 400 MHz 'Benelux Cross Project'. Here he worked with the Swedish astronomer Jan Högbom, who had finished a PhD with Martin Ryle at Cambridge in August 1959 on 'The Structure and Magnetic Field of the Sun', his thesis including chapters on aperture synthesis theory and earth rotational synthesis.) When the Belgians pulled out, the Benelux project was abandoned for what ultimately became the Westerbork Synthesis Telescope. Chris maintained an active collaboration with the Dutch group developing the new telescope. Jan Högbom also played a key role. The book *Radiotelescopes* that he and Chris wrote together was published by Cambridge University Press in 1969 (39). A Russian translation by Yuri Ilyasov appeared in 1971 (41) and a Chinese translation by Chen Jian-sheng in 1977 (40).

Back in Sydney, Chris attempted, unsuccessfully, to gain financial support to build a 'hole-in-the-ground' spherical reflecting telescope 30m in diameter for use at millimetre wavelengths.

Chris was keen to have a big project that could be used to stretch the minds of postgraduate students. Undeterred by the failure of the 'hole in the ground' proposal, he continued looking for ways forward. After hearing from Paul Wild that the field station at Fleurs was to be 'bull-dozed', he approached CSIRO and requested that CSIRO donate his grating cross area at Fleurs to the University of Sydney. As he said later: 'CSIRO was very generous. We took over the whole field station.'

In the period since Chris had taken up his appointment at the University of Sydney, Martin Ryle at Cambridge had demonstrated a full rotational synthesis system and embarked on the design of the Cambridge One Mile Telescope, while the Dutch had decided to proceed with a synthesis telescope at Westerbork in place of the original Cross concepts.

Chris saw the opportunity for a significant and challenging project and set about changing the arms of the 'Chris Cross' instrument to convert it from one of low sensitivity and 3 arc min resolving power, designed for solar observations, into a high-resolution rotational synthesis telescope at roughly the same wavelength of 21cm as the previous Fleurs crossed multi-element interferometer. As the Fleurs Synthesis Telescope, it would be used for galactic and extra-galactic astronomical observations in the years 1973-1988.

The Fleurs Synthesis Telescope (FST) (42) was built utilizing the infrastructure of the Chris Cross. The compound interferometer at 1415 MHz or 21 cm was formed by multiplying each of 5.8 metre elements of the original Chris Cross by two new 13.7 m antennas on the E-W and N-S arms with a total baseline of 800 m. Astronomical sources could be tracked for eight hours per day. With this combination, roughly circular beams could be achieved from the celestial equator to a declination of -80d. The resolution was 40 arc sec after two times eight hours, once with the E-W array and once with the N-S array. During its years of operation, the FST provided the highest-resolution radio telescope (apart from long-baseline Interferometry) in the southern hemisphere. The FST is shown in Figure 4.

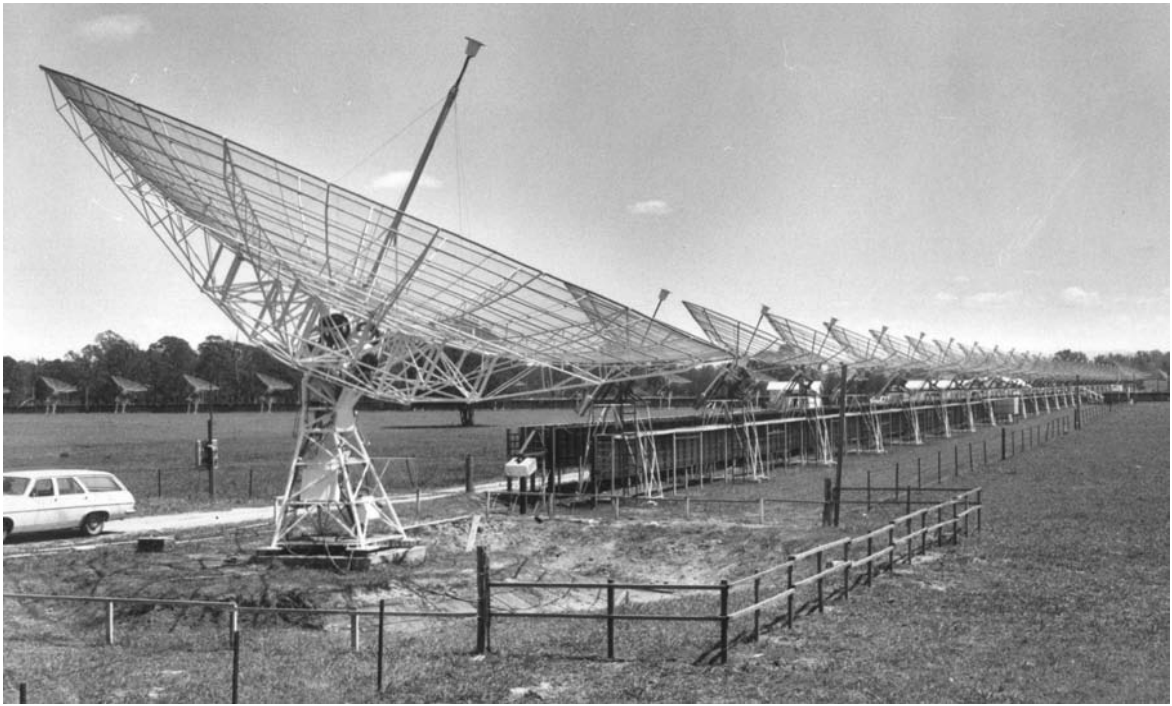

Figure 4. A view along the array of the Fleurs Synthesis Telescope in1973

(ATNF, Historic Photographic Archive, 9097-11).

Over a period approaching 25 years, the FST was an important test-bed for PhDs in Electrical Engineering and a training ground for numerous scientists who have played major roles not just in radio astronomy but in Australian industry, universities and the CSIRO. Chris had a strong belief in the benefit for students of working on large complex projects in a team environment, and their subsequent success is a significant tribute to him. During the life-time of the FST, a number of technical improvements were made that expanded the capabilities of the instrument. Its sensitivity was improved in late 1975 by about a factor of 4 with the addition of more sensitive receivers to the 13.7 m antennas. Starting in 1976, plans were being made to extend the FST to a longer baseline and hence greater angular resolution (51). By 1984, the resolution of the array was doubled to about 20 arc sec by the addition of two new 13.7 metre antennas.

During this period, the FST produced numerous radio images and astrometry at the arc sec level; the success of these research endeavours was a major stimulus to the planning of the Australia Telescope (AT) that bgan in the early 1980s. The FST was a 'hands on test-bed' for many of the concepts that were later used at the AT. An additional indirect influence was the pronounced impact of the FST on the Australian astronomical community; the breakthroughs brought about by high-sensitivity, high-resolution radio images showed the need for a high-resolution multi-frequency instrument.

Chris created a unique environment at the FST, characterized by a symbiotic relation between young innovative electrical engineers and radio astronomers; the two groups were learning from each other as technical innovations led to challenging radio astronomical observations. A prominent example was the solution of the projection of the three dimensional uv space (the Fourier components of the observed radiation field) into a two-dimensional co-ordinate system (Frater and Docherty, 1980).

Starting in 1975, the FST was in full operation at 21 cm in the continuum , with a large number of observations completed with the E-W arm. In late 1975, the N-S arm came into operation. For most fields the final observational product consisted of two eight-hour observations with the EW and NS arms. Circular beams were possible over most of the southern sky. For the next thirteen years, the FST was used extensively by more than 75 scientists to publish 69 papers.

It is appropriate here to provide a short summary of a number of highlights of the early years of FST research (roughly up to the time of Chris Christiansen's retirement from the University of Sydney in 1979). A further paper we are preparing will provide a detailed discussion of the scientific impact of the FST, including a complete bibliography of FST publications.

In an initial publication from 1976, Frater et al. described a calibrator grid of compact radio sources. This set of calibration positions for the southern sky was established by observations of four sources of previously known positions over the declination range  -75 to -13 d.  These observations were necessary in order to carry out routine observations over most of the southern sky at declinations outside the ranges of the major high-resolution radio telescopes in the northern hemisphere.

A keynote publication of this era was a 1977 paper by Christiansen et al. (48); this highlighted the capabilities of the FST. The paper included images of the famous radio galaxies Centaurus A, Pictor A, PKS 0349-27.9 and  the low galactic latitude source G309.7+1.7, a likely extragalactic source close to the galactic plane. Christiansen was also a co-author in a 1977 publication by Goss et al. (47).

In 1976 the astronomical community was excited by the discovery of the second optical pulsar, following the discovery of the Crab pulsar seven years earlier. The newly completed high-resolution FST played a key role in this discovery by the determination of the sky co-ordinates at 1415 MHz (observations 4 to 6 September 1976). The Molonglo Radio Telescope was also used in this determination. The accuracy of the combined position was about 1 arc sec (Goss, Manchester, Frater and McAdam, 1977).  A few months later (24-27 January 1977) a group of astronomers (including Goss and Manchester) detected  optical pulses from the 23.7 mag star M ( of Lasker 1976) using the Anglo-Australian Telescope based on the new radio position. The paper was published in *Nature* on 21 April 1977 by P. T.  Wallace and eleven co-authors.

The detection of the second optical pulsar had a major impact throughout the astronomical world. John Bolton wrote on 3 May 1977 to his former Chief of the Division of Radiophysics, Taffy Bowen, who was in the USA at this time (National Archives of Australia C 4661): 'A combined effort from Greenwich, the AO [AAT] and RP [Radiophysics] recently found the optical pulsar from the Vela pulsar in the position resulting from a combined effort at Molonglo and Fleurs. It's 1999,000 times weaker than the Crab (in the optical).'

In the late 1970s, an important new component of the FST research programme was initiated, the study of galactic supernovae remnants (SNR). In 1977, Ian Lockhart et al. published a paper providing the first high-resolution image of the unusual SNR G282.0+1.8 at 1415 GHz. They suggested that the entire source was a pulsar-powered wind nebula emitting synchrotron radiation. This remarkable SNR is a non-shell type SNR; the FST image led to an optical identification by Goss et al. (1979), showing this to be an oxygen-rich SNR. The optical spectrum exhibited prominent [OIII] and [OI] with no prominent ☐☐ emission. The modern history of the source is summarized by Gaensler and Wallace (2003), who used data from the Australia Telescope Compact Array to show that the central source is a pulsar-powered wind nebula with a flat spectral index; the source is powered by the young pulsar J1124-5916 (2900 years characteristic age). The source is surrounded by a steeper radio spectrum outer plateau. Remarkably, the ATCA (at 20, 13 and 6 cm) images are quite similar to the FST 20 arc sec image made in the mid-1980s by Milne et al. (1985) using the expanded FST. In this FST image, the pulsar wind nebula is clearly delineated from the SNR shell plateau. (This image was presented at the Alec Little Memorial Symposium of the Astronomical Society of Australia in late 1985.)

In the period 1979-1987, twelve additional publications appeared with descriptions of the 1415 MHz properties of 25 SNR in the Galaxy. Jim Caswell led a number of colleagues in this collaboration. A wide variety of structures in these SNR were observed; the classification into the two classes of shell and filled-centre SNRs remains a useful paradigm today. A noteworthy image of the radio remnant of AD 1006 (G327.6+14.6) was obtained by Caswell et al. (1983); with a resolution of about 1 arc min, the 28 arc min diameter source was well  resolved. Detailed comparison with optical and X-ray data showed that the images at the various wavelengths showed a general but no detailed correspondence.

## Air Navigation Work

In addition to the work in radio astronomy, and as a further step in establishing challenging projects for postgraduate work, Chris established a research group to tackle problems in the radio navigation of civil aircraft, initially led by Robert Redlich.  It was later led by Godfrey Lucas, who with his group developed significant improvements to the Instrument Landing System (ILS) and Vector Omnirange System (VOR). This group earned a considerable reputation internationally.

# Links with China

Chris's interest in China went back to his childhood:

> My interest started in about 1920 through an aunt who was matron of a missionary hospital in China. I later read a book called Red Star over China, by Edgar Snow, and became still more interested in China. When I had to go to the IAU meeting in Japan in 1961 I thought, 'Why don't I try to get into China?' So I wrote to the Chinese Academy, saying that I was representing our Academy, more or less, at the meeting and asking if they could help me to get a visa to China. Nothing happened for quite a while, and then suddenly they said they would be glad for me to go as their guest, and give lectures and so on. They'd read everything I had ever written, I think. After that the connection just increased, because while I was there I met the President of their Academy and I suggested that he ask our Academy to send a delegation to China. A delegation did go, probably in 1964, and in return a Chinese one came to Australia.

Chris first visited China in 1963. At that stage, an embryonic radio astronomy group was struggling with some copies of Russian solar radiometers. The group was interested in building a solar array but lacked simple things like cables. Chris was able to draw on his earlier experience to help them with open-wire transmission lines. He formed a strong friendship with Wang Shouguan. Wang and Wu Huai-wei visited Sydney in 1964. The group at Fleurs collaborated in the development of the new telescope at Miyun Observatory. Chris made many visits to China over the years, while Chen Hong-shen and Ren Fang-bin visited Fleurs in 1974. Chris organized lecture tours and visits to the Miyun group by Bob Frater in 1976 and Miller Goss in 1977 and Chin Kwong visited China in 1979 as part of that collaboration. Chen Jian-sheng and colleagues visited Australia in 1979 to observe at Fleurs (Chen Jian-sheng et al., 1982).

Wang Shouguan has written up the story of the relationship (Wang Shouguan, 2009). In recognition of his 'long and important contribution to Chinese astronomy', Chris was elected a Foreign Member of the Chinese Academy of Sciences in 1996.

# Local and International Roles in Science

Chris took on significant roles with the international scientific unions. His involvements with both the International Astronomical Union and the International Union for Radio Science began in the 1950s. He served as a Vice-President of the International Astronomical Union (IAU) from 1964 to 1970, President of the Radio astronomy Commission of the International Union for Radio Science (URSI) from 1963 to 1966, Vice-President of URSI 1972 to 1978 and then President from 1978 to 1981. He was appointed a Life Honorary President of URSI in 1984. He was a member of the General Committee of the International Council of Scientific Unions (ICSU) from 1978 to 1981.

In the Australian Academy of Science, to which he was elected in 1959, he was Foreign Secretary 1981-85, served on the Council for two terms, and was chairman of several committees including the National Committee for Radio Science, 1962-70. He was President of the Astronomical Society of Australia, 1977-79.

He was a member of the Australia-China Council of Australia's Department of Foreign Affairs, 1979-82. He was a UNESCO Consultant in India on the construction of a 'Giant Equatorial Radiotelescope' in the 1980s.

Chris had extensive scientific linkages, both local and international. He was always looking to involve young engineers and scientists in the work of international bodies, leading to one of us (RHF) finding himself, at the age of 32, as the Commission D representative for Australia on URSI. Chris actively recruited young representatives to Academy committees and the like.

His long involvement with both the IAU and URSI, together with his many interactions with peers around the world, gave him an extraordinary international network. This came into play over the years in arranging visits, finding post-doctoral positions and so on.

Chris Christiansen died on 26 April 2007. He is survived by his sons, Steve and Tim. His wife, Elspeth, died in 2001, while their son Peter, a well-known atmospheric scientist at the University of Sussex, died in 1992.

**The Legacy**

Those involved with the Fleurs Synthesis Telescope, as staff members as Bob Frater was or as postgraduate students, learned a lot from the process of developing the instrumentation necessary for an operational telescope. When Frater went on to run the Division of Radiophysics and to build the Australia Telescope, these people played an essential role, and even today they are playing key roles in the developmental work for the Square Kilometre Array. The success here is part of Chris's legacy.

The challenge of the work at Fleurs and the team environment created by Chris yielded other benefits that also stand as his legacy. Key parts of the 802.11 wireless LAN work that led to the formation of the company Radiata, that has delivered such handsome royalties to CSIRO, was done by people – John O'Sullivan, Graham Daniels, Terry Percival in the patent process and David Skellern in the commercialization who went through this system as students. Following the success of this project, Bob Frater prepared a presentation titled 'Rome Wasn't Built in a Day' to emphasise the linkages back to Chris's Fleurs vision and the many other significant outcomes that have resulted. This is written up in Matthews and Frater (2003 and 2008).

In addition to his significant influence in the development of the Westerbork Synthesis Telescope, Chris's influence on the development of radio astronomy in both India (with the links to Govind Swarup) and China (with Wang Shouguan) is inestimable.

Chris truly carried the high status that Australia enjoyed in the early days of radio astronomy through to the present day through the training and experience that came out of Fleurs. In addition to the PhDs in Electrical Engineering, a smaller number of PhD degrees were awarded in astronomy through the University of Sydney's School of Physics and the Australian National University. Numerous collaborations were established between the University of Sydney (Electrical Engineering and the School of Physics) and other astronomical groups in Australia (CSIRO Division of Radiophysics, Australian National University and the Anglo-Australian Observatory), associations that have continued well into the new century.

Chris' astronomical legacy consists of two major parts:

(1). Major instrumental innovations that had an impact on radio astronomy. Earth rotational synthesis was achieved in 1955 with the Potts Hill grating array. A few years later, the crossed-grating multi-element interferometer at Fleurs was completed, an instrument based on the twin concepts of the grating array as had been used at Potts Hill and the Mills Cross technique of correlating two orthogonal arrays. After Chris's move to the University of Sydney in 1960, the establishment of a remarkable group in Electrical Engineering led to the Fleurs Synthesis Telescope.

(2). New astronomical understanding:

(a). The pioneering HI line work in the first years of 21 cm hydrogen spectroscopy, starting in 1951. This work showed the existence of spiral arms in the gaseous component of the Milky Way; the southern Galaxy was imaged contemporaneously with imaging of the northern Milky Way by the Leiden group.

(b). The determination of the properties of the decimetre Sun in the early 1950s at 21 cm (1.4 GHz). At 21 cm, the radiation arises from the transition region between the corona and the outer chromosphere. In this region the change-over between the steady optical Sun and the spectacularly variable metre-wave Sun occurs. Thus it was possible to determine physical conditions in the most important region of solar activity. The observations were carried out at Potts Hill using the EW and NS grating arrays from 1952 and at Fleurs using the crossed multi-element interferometer starting in 1957, on the first day of the International Geophysical Year. The quiet Sun, arising from thermal free-free emission from the solar atmosphere, was investigated in detail, with the detection of the prominent equatorial limb brightening at solar minimum using Potts Hill data from 1953. With the Fleurs array, two-dimensional images were created by scanning the sun during an observations of about an hour; the image, with a resolution of about 3 arc min, was constructed 'television fashion' by using scans at different latitudes on the sun. With this method, detailed images of the 'slowly varying component' were obtained at a rate of one image per hour. (The term 'slowly varying component' was probably invented at Radiophysics in the early 1950s; Christiansen and Pawsey championed the use of the term, which became the standard terminology within a few years.). The electron density and temperature (~3e9 cm-3, ~2e6 K) of the radio 'plages' were determined at elevations of ~ 20,000 km above the Sun's surface; the typical size of the radio plagues was observed to be about 300,000 km. The 'slowly varying component' arises from a combination of free-free emission and gyro resonance radiation in localized regions of higher electron density and magnetic field at locations above sunspots or chromospheric plages. Since the temperature of the radio plages was found to be comparable to the corona, the regions of enhanced emission at 21 cm can be associated with distinct regions of increased emission that were optically thick at 21cm. A few years later, during the solar maximum of 1958, the quiet Sun was imaged by Labrum (1960) using the Fleurs array; he confirmed the shape of the earlier work from Potts Hill and found that the brightness of the quiet-sun emission at the solar maximum was about twice that of the earlier data from Christiansen and Warburton, obtained during the previous solar minimum.

## Awards

Chris was elected a Fellow of the Australian Academy of Science in 1959, an Honorary Fellow of the Institution of Engineers, Australia, and a Fellow of the Institution of Electrical Engineers (UK), the Institution of Radio and Electronic Engineers (Aust), the Institute of Physics (UK), the Australian Institute of Physics, and the Royal Astronomical Society (UK).

On retirement in 1979 from the Chair of Electrical Engineering at the University of Sydney, Chris was given the title of Professor Emeritus and later awarded the degree of Doctor of Science in Engineering, honoris causa.

He was a Visiting Fellow at the Mount Stromlo Observatory of the Australian National University from 1980 to 1985.

He also received honorary degrees from the University of Melbourne (1982) and the University of Western Sydney (1994) and was elected a Fellow of the Chinese Academy of Sciences in 1996.

He was awarded the Fleming Premium of the Institution of Electrical Engineers in 1961, the Peter Nicol Russell Medal, the premier award of the Institution of Engineers Australia, in 1970, and the medal of ADION from the Nice Observatory in France, in 1976.

## Acknowledgements

This memoir is based on a number of interviews of Chris Christiansen (Sullivan [1976], Bhathal [1996], Crompton [1997]), his publications, published historical material (Orchiston et al. 2006, Wendt et al. 2008a, b, c), the authors' personal knowledge from the early 1960s until Chris's death, and discussions with many people who worked at Fleurs Observatory over the years. Harry Wendt and Stephen White provided valuable input. Tim Christiansen reviewed the manuscript and provided useful feedback. The National Radio Astronomy Observatory is operated by Associated Universities, Inc. under cooperative agreement with the National Science Foundation.